\documentstyle[12pt]{article}
\newlength{\abstractwidth}
\begin{document}
\thispagestyle{empty}
\pagestyle{plain}
\input{epsf}
%%%%%%%%%%%%%%%%%
%macros
\def\beq{\begin{eqnarray}}
\def\eeq{\end{eqnarray}}
\def\nn{\nonumber\\}
%%%%%%%%%%%%%%%%%%%%
\renewcommand{\thefootnote}{\fnsymbol{footnote}}
\renewcommand{\thanks}[1]{\footnote{#1}} % Use this for footnotes
\newcommand{\starttext}{
\setcounter{footnote}{0}
\renewcommand{\thefootnote}{\arabic{footnote}}}
%%%%%%%%%%%%%%%%%%%%%%%%%%%%%%%%%%%%%%%%%%%%%%%
\begin{titlepage}
\bigskip
\hskip 3.7in\vbox{\baselineskip12pt
\hbox{CERN-TH/99-211}\hbox{hep-th/9908024}}
\bigskip\bigskip\bigskip\bigskip

\centerline{\large \bf BACK-REACTION TO DILATON-DRIVEN INFLATION}

\bigskip\bigskip
\bigskip\bigskip

\centerline{\bf A. Ghosh, R. Madden, G. Veneziano\thanks{amit.ghosh,
richard.madden, gabriele.veneziano@cern.ch}}
\medskip
\centerline{CERN, Theory Division}
\centerline{CH-1211, Geneva 23}

\bigskip\bigskip

\begin{abstract}
\baselineskip=16pt
%%%%%%%%%%%%%%%%%%
We compute the leading-order back-reaction to dilaton-driven
inflation, due to
graviton, dilaton and gauge-boson production. 
The one-loop effect turns out to be
non-vanishing (unlike
 the case for pure de-Sitter and for power-law inflation), to
be of relative order $\ell_P^2H^2(t)$, and to have the correct sign
for favouring the exit to a FRW phase.
%%%%%%%%%%%%%%
\end{abstract}
\vspace{3cm}
\begin{flushleft}
CERN-TH/99-211\\
\today
\end{flushleft}
\end{titlepage}
%%%%%%%%%%%%%%%
\starttext
\baselineskip=18pt
\setcounter{footnote}{0}
%%%%%%%%%%%%%%%%%%%%%%%%%%%%%%%%%%%%%%%%%%%%%%%%%%%%%%%%%%%%%%%%
\section{Introduction}

In recent years, pre-big bang (PBB) cosmology \cite{PBB}
has established itself as a  logically consistent alternative to
standard slow-roll inflation \cite{inflation}.
While the ultimate verdict on which, if either, of these two versions  
of cosmology will survive is left to future experiments,
at the theoretical level
each one has its own advantages and shortcomings.

 The trade-off follows from the fundamental physical
difference between the two scenarios: in slow-roll inflation the
evolution is
 from larger to smaller
curvatures, while in  dilaton-driven PBB inflation the opposite is true.
 Thus, while in  the former case
Planckian/string-scale
curvatures are naturally reached at the very beginning, pushing the
problem of
initial conditions into the full quantum-gravity domain (see e.g.
\cite{HTL}),
in the latter
that problem can be tackled  within a perfectly known (albeit
non-linear) set of equations (see e.g. \cite{initial}). On the other  
hand, since
large curvatures naturally occur
 at the end of dilaton-driven
inflation\thanks{A  phenomenologically interesting consequence
of this difference
is that,  in slow-roll inflation, Planck/string-scale physics
 is ``screened" from observation by
the subsequent long inflationary era; in the PBB scenario,
it is in principle accessible to observation.}, the exit problem becomes
harder to analyse than in the conventional scenario.

It is generally believed that the exit problem of string cosmology
\cite{exit}, i.e.
achieving a smooth, non-singular transition from inflationary to  FRW-type
string cosmology solutions,
should be solved by a combination of two kinds of effects,
corresponding to two different classes of corrections
to the low-energy, tree-level effective action.
While higher-derivative corrections  should be able to
bound the growth of curvature \cite{alpha'} (or to allow a
reinterpretation of the
lowest-order curvature
singularity), loop corrections \cite{loops}, accounting for the
back-reaction on the geometry from
particle production, should  stop the growth of the
coupling (of the dilaton) when a certain critical-energy condition
\cite{critical}
 is reached.
Recent proposals of cosmological entropy bounds \cite{entropy}
appear to support \cite{support} these ideas.

Unfortunately, present techniques are still inadequate for giving a
convincing answer
to the question of whether any one of these desired effects does
actually take place.
 The question of fully non-perturbative higher-derivative corrections  
amounts to
finding an exact (2D) conformal field theory. However, since
supersymmetry is spontaneously broken in the case of cosmology, chances 
to get such a CFT in closed form are slim.

Similarly, understanding the full effect of back-reaction would require
knowledge of the full one-loop effective action up to four derivatives.
In spite of brilliant work by Vilkovisky and collaborators \cite{Vilko}
on this problem, these are still early times for applying
the results so far obtained to our
problem. Thus our aim here will be more modest: we will try to
understand in which direction does the back-reaction from particle
production work in
PBB cosmology while it is still a small effect. No attempt will be
therefore be made to achieve (a fortiori to describe) a complete exit.

We thus took an easier road, recently opened by Iliopoulos et al.
\cite{Iliop} and by
Abramo and Woodard \cite{AW1}, \cite{AW2}. Their method was already
checked,
whenever possible, against the
more intuitive approach of Abramo, Brandenberger and Mukhanov  
\cite{Brandem}.
Apart from our use of a different gauge, known to be
more appropriate in string cosmology, our work can be seen as a
straightforward application
of that method in the context of string cosmology. As a check of
gauge invariance, we will also reproduce some
results of \cite{AW1}, \cite{AW2}.

Another (apparently strong) limitation of our work is that
we consider loop corrections
to solutions of the low-energy effective action,
i.e. we do not include higher-derivative terms
in the latter. Since the back-reaction effect
turns out to be of the same order as a four-derivative
term, it may look suspicious to neglect such terms in the tree-level
action while keeping
them in the correction.
There is, however, a limit in which such an approximation is justified:
 it is a large-$N$
limit, where $N$ is the number of species of the produced particles.
The one-loop back-reaction is of relative order $g^2 N\ell_s^2 R$
and thus dominates over
a tree-level higher-derivative correction ${\cal O}(\ell_s^2 R)$ at
sufficiently
 large\thanks{$N$
is indeed a large number for a typical
string theory gauge group, say $E_8 \otimes E_8$.} $g^2N$.
Finally, it can also be shown that higher-loop effects on the metric,  
the dilaton and the
antisymmetric tensor (axion) are parametrically smaller at large $g^2  
N$ and small
$g^2$,  justifying the one-loop approximation. This observation makes  
it all the more urgent to devote further
effort to a non-perturbative understanding of
one-loop back-reaction in cosmology.

The paper is organized as follows: after recalling (Sections 2 and 3)
the form of the lowest-order PBB background whose modification we wish 
to compute, and the
general idea of the method, we will introduce (Section 4) our gauge
choice and then go, in
Section 5,
to the heart
 of the calculation of the back-reaction, first from gravi-dilaton  
production,
 and then from gauge-bosons production.
 In Section 6, after expressing the final result in a convenient
form, we will argue that
the back-reaction is of the expected order of magnitude and that,
at leading order at least, it modifies the original
background in the right direction to help with the exit problem.
Finally, an explicit check of the gauge independence of our results  
is presented in an Appendix.

\section{PBB Backgrounds and Fluctuations}
\setcounter{equation}{0}

In units in which $16 \pi G = \hbar = c =1$ the
normalized four-dimensional gravi-dilaton effective action in the
Einstein frame reads:
 \beq
S=\int d^4x\sqrt{-g}\left[R-{1\over 2}(\nabla\varphi)^2\right]\;.
\label{action}
\eeq
Even within this minimal set of moduli (corresponding to having
frozen the axion as well as internal dimensions),
 the gravi-dilaton system offers several
interesting cosmological solutions, including, of course, the simplest ones,
which describe spatially flat, homogeneous and isotropic Universes.
In the comoving (cosmic)-time frame the  line element takes the
form
\beq
ds^2=-dt^2_0+a_0^2(t_0)\,d{\boldmath \vec x}\cdot d{\boldmath \vec x}\;,
\label{cometric}\eeq
where $t_0$ represents comoving time (we shall often use the suffix $0$
to denote  tree-level quantities).  In the discussion of
quantum fluctuations it is often convenient to go over to the
so-called conformal frame, in which the metric (\ref{cometric}) becomes
\beq
ds^2=\Omega^2(\eta)\left(-d\eta^2+d{\boldmath \vec x}\cdot
d{\boldmath \vec x}
\right) \; .
\eeq
We recall that comoving and conformal times are related  by
$dt_0=\Omega d\eta$\,.
   In the conformal
frame a typical PBB-type solution appears as follows:
\beq
\Omega=\left({\eta\over\eta_{\,i}}\right)^{1/2},\qquad
\varphi_0=-\sqrt 3\ln\left(-\eta\right)\,,\qquad\eta_{\,i}<\eta<0\;,
\label{solution}\eeq
where, by identifying the (arbitrary) constant
 $\eta_{\,i}$ with the beginning of dilaton-driven inflation (DDI),
we have normalized $\Omega$ to $1$ at that moment.
It is convenient to define, as usual,  ${\cal
H}(\eta)=\Omega'/\Omega=1/2\eta$\,, which is
related to the  physical Hubble parameter $H$ by
\beq
H_0={\dot a_0\over a_0}={\Omega'\over\Omega^2}=H_i\Omega^{-3}
={1\over 3t_0}\,,\qquad t_0<0\;,\label{zerohubble}
\eeq
where $H_i=1/2\eta_{\,i}$ refers to the initial value of the physical
Hubble parameter. Note that, at initial time $\eta=\eta_i$, the physical
Hubble parameter is given by $\Omega'$ since $\Omega_i=1$.

We remind the reader that the above background corresponds to
DDI in the string frame, with a string-frame scale factor blowing up
like $(-t_s)^{-1/\sqrt3}$ when the comoving string time $t_s$ approaches
$0$ from negative values $(t_s\to 0^-)$. In the Einstein frame, this  
is seen as a
contraction
(gravitational collapse, see \cite{initial}). What remains frame-independent
is the fact that curvature grows in time, rather than decreases as in
power-law inflation. This is why the results of \cite{AW2} cannot be  
directly used
in our context.

Let us now define the fluctuations around the classical backgrounds
by:
\beq
g_{\mu\nu}=\Omega^2\left(\eta_{\mu\nu}+\psi_{\mu\nu}\right)\,,\qquad
\varphi=\varphi_0+\phi\;.\label{fluc}
\eeq
Our goal here is to  compute the back-reaction on the above homogeneous
backgrounds (\ref{solution}) (i.e. $\psi_{\mu\nu}$ and $\phi$) due to
 quantum fluctuations at leading order.
Following \cite{AW1}, \cite{AW2}, we will thus parametrize
the correction to the  solution (\ref{solution}) by
\beq
\langle\psi_{\mu\nu}\rangle&=&A(\eta)\bar\eta_{\mu\nu}+C(\eta)t_\mu
t_\nu\nn
\langle\phi\rangle&=&D(\eta) ~~ ,
\label{acd}
\eeq
where $t_\mu\equiv(-1,0,0,0)$ is a time-like vector and $\bar\eta_{\mu\nu}
=\eta_{\mu\nu}+t_\mu t_\nu\equiv{\rm diag}\,(0,1,1,1)$\,. In the next
section we shall recall the method of Refs. \cite{Iliop}, \cite{AW1},  
\cite{AW2}
for  computing  $A,C$ and $D$. Here we just conclude by
recalling that $A,C$ and $D$ have no physical meaning separately.
For instance, if $A$ and $D$  are given in terms of $C$ by
\beq
A &=& {{\cal H}_0 \over \Omega}(\eta) \int  ^{\eta} d \eta' (\Omega  
C)(\eta') \nn
D &=&  {1 \over 2} {\varphi_0' \over \Omega}(\eta)
 \int  ^{\eta} d \eta' (\Omega C)(\eta') ~~,
\eeq
then all three can be gauged away by the redefinition of time, $\eta  
\to \eta +
{1 \over 2} \Omega^{-1} \int  ^{\eta}  \Omega C$.

\section{The Procedure}
\setcounter{equation}{0}

Since our background is  of the power-law (though not of the
power-inflation)
type, we shall closely follow here the procedure described in \cite{AW2} 
for computing $A,C,D$. We will then stress at which point
the two procedures start to diverge.
 In \cite{AW2} the authors first compute some amputated
one-point functions denoted by $\alpha, \gamma$ and $\delta$ (see below),
 and then
attach external leg propagators in order to reconstruct the  desired
functions $A,C,D$.
Let us describe the procedure in a different, perhaps simpler, way.

If, generically, we wish to compute  a fluctuation
 $\Psi_i$  over a generic
background $\Psi_{0i}$, then the relevant one-point function
is given by the obvious functional integral
\beq
\langle\Psi_i(x)\rangle={\int D\Psi_\ell e^{iS[\Psi_{0\ell}+\Psi_\ell]}
\Psi_i(x)\over\int D\Psi_\ell e^{iS[\Psi_{0\ell}+\Psi_\ell]}}\;.
\eeq
If the action is expanded only up to the quadratic order in
$\Psi_\ell$, the
one-point functions are trivially zero. If, however, the action is
expanded up to cubic order,
$S[\Psi_{0\ell}+\Psi_\ell]=S[\Psi_{0\ell}]+S_2^{ij}
[\Psi_{0\ell}]\Psi_i\Psi_j+S_3^{ijk}[\Psi_{0\ell}]\Psi_i\Psi_j\Psi_k$
(the summation
convention is adopted everywhere), and
$S_3^{ijk}[\Psi_{0\ell}]$ is treated perturbatively, then there
is a non-trivial one-loop correction to the one-point functions (see  
Fig. 1).
This is the familiar story of vacuum-shifting tadpoles in quantum
field theory. Analytically the result can be expressed as follows:
\beq
\langle\Psi_i(x)\rangle={\int D\Psi_\ell e^{iS[\Psi_{0\ell}]+iS_2^{ij}
[\Psi_{0\ell}]\Psi_i\Psi_j}iS_3^{jkn}[\Psi_{0\ell}]\Psi_i(x)\Psi_j\Psi_k\Psi_n
+...\over\int D\Psi_\ell
e^{iS[\Psi_{0\ell}]+iS_2^{ij}[\Psi_{0\ell}]\Psi_i\Psi_j}}
\;.\label{loops}
\eeq
The calculation of the diagrams in Fig. 1
\begin{figure}[htb]
\epsfxsize=1.7in
\centerline{\epsffile{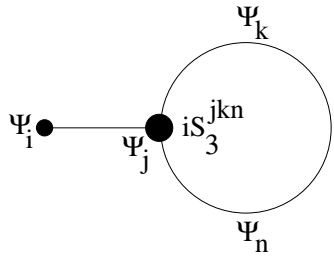}}
\caption{\em\baselineskip=11pt The one-point functions from tadpoles}
\end{figure}
can be performed in two
steps. First, we compute the amputated one-point function
 for all possible three-point
vertices,
$iS_3^{jkn}$. Then the full one-point functions are obtained by
attaching the
appropriate external lines. Formula (\ref{loops}) gives
\beq
\langle\Psi_i(x)\rangle=\int d^4z\langle\Psi_i(x)\Psi_j(z)\rangle
\langle\Psi_k(z)\Psi_n(z)\rangle[iS_3^{jkn}]\;,\label{define}
\eeq
where (dropping indices hereafter),
$\langle\Psi(x)\Psi(z)\rangle$ is the standard Feynman propagator
 for  the quadratic action $S_2[\Psi_0]\Psi^2
\equiv\Psi\Delta\Psi$, i.e.
\beq
\Delta(x)\langle\Psi(x)\Psi(z)\rangle=i\delta^4(x-z)\;.
\eeq
Finally, one can  express the one-point function as:
\beq
\langle\Psi(x)\rangle=-\Delta^{-1}(x)\langle\Psi(x)\Psi(x)\rangle
[S_3]\;.\label{final}
\eeq
It turns out that the coincident limit propagator $\langle\Psi(x)
\Psi(x)\rangle$ is a function of time, $x^0$, only. Therefore, if one  
can invert the
kinetic operator in the zero-momentum limit, then  (\ref{final})
provides our final expression for the one-point function. For the purpose of
the first step it is convenient
to distinguish three kinds of amputated one-point functions,
$\alpha$, $\gamma$ and $\delta$, defined respectively in Figs. 2, 3 and 4.

As an illustration of the use of (\ref{final}), we present here
the calculation of the one-point function
$A$ in the covariant gauge introduced in \cite{Iliop}. Note, from
(\ref{acd}), that  $A$ is nothing but the
amplitude $\langle\psi_{11}\rangle$, which can be expanded
from the general definition (\ref{define}) and Fig. 1, in terms of  
the amputated
one-point functions $\alpha, \gamma$ and $\delta$ as:
\beq
A\to\langle\psi_{11}\phi\rangle\cdot\delta+\langle\psi_{11}\psi_{\mu\nu}\rangle  
t^\mu t^\nu\cdot\gamma+\langle\psi_{11}\psi_{\mu\nu}\rangle\bar
\eta^{\mu\nu}\cdot\alpha\;.\label{A1}
\eeq
As in (\ref{final}), the external line propagators  can be
replaced by the inverse of the kinetic operators in the zero-momentum limit.
 In the covariant gauge of Iliopoulos
et al.  the Lagrangian can be diagonalized by transforming the variables
into suitable linear combinations.
 The propagators of the original variables can be read
off from those of the diagonalized variables by applying
 the inverse transformation.
For a standard power-law-inflation background, $\Omega\sim t^s$ with
$s>1$, one easily finds
\beq
\langle\psi_{11}\phi\rangle&=&-{\sqrt s\over s+1}i\Delta_A+{\sqrt
s\over s+1}
i\Delta_C\nn
\langle\psi_{11}\psi_{\mu\nu}\rangle t^\mu t^\nu&=&{1\over
s+1}i\Delta_A+{s\over s+1}
i\Delta_C\nn
\langle\psi_{11}\psi_{\mu\nu}\rangle\bar\eta^{\mu\nu}&=&\left(-4+{3\over
s+1}\right)i\Delta_A
+{3s\over s+1}i\Delta_C\label{recipe}
\eeq
where $i \Delta_A$ and $i \Delta_C$ are particular propagators of the  
diagonalizing
variables. They can be replaced by the inverses of the
corresponding kinetic operators in the zero-momentum limit,
$1/D_A$ and $1/D_C$.
With this replacement, when (\ref{recipe}) is
plugged back into the expression for the one-point function $A$ in
(\ref{A1}),
we immediately reach
the desired expression for $A$:
\beq
A={1\over D_A}\left[-4\alpha+{1\over s+1}(3\alpha+\gamma)-{\sqrt s\over s+1}
\delta\right]+{1\over D_C}\left[{s\over s+1}(3\alpha+\gamma)+{\sqrt  
s\over s+1}
\delta\right]\;.\label{A2}
\eeq
This equation is the same as was given in \cite{AW2}. The above outline shows
how simply the relevant formulae can be obtained
once the general result (\ref{final}) has been derived.
In the Iliopoulos et al. gauge, which does need ghosts,
there are three different types of loops, shown in Figs. 2, 3, and 4, where
the dotted line describes the
graviton, the dashed line the ghost, and the solid line
the dilaton.
\begin{figure}[htb]
\epsfxsize=3in
\bigskip
\centerline{\epsffile{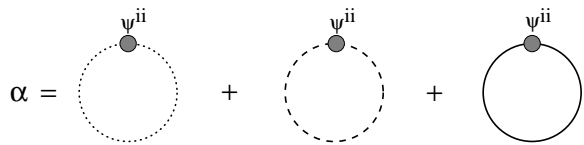}}
\caption{\it\baselineskip=12pt
The $\alpha$-type amputated one-point function}
\bigskip
\label{alpha}\end{figure}
\begin{figure}[htb]
\epsfxsize=3in
\bigskip
\centerline{\epsffile{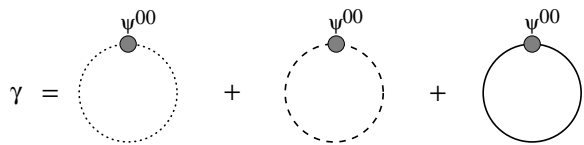}}
\caption{\it\baselineskip=12pt
The $\gamma$-type amputated one-point function}
\bigskip
\label{beta}\end{figure}
\begin{figure}[htb]
\epsfxsize=3in
\bigskip
\centerline{\epsffile{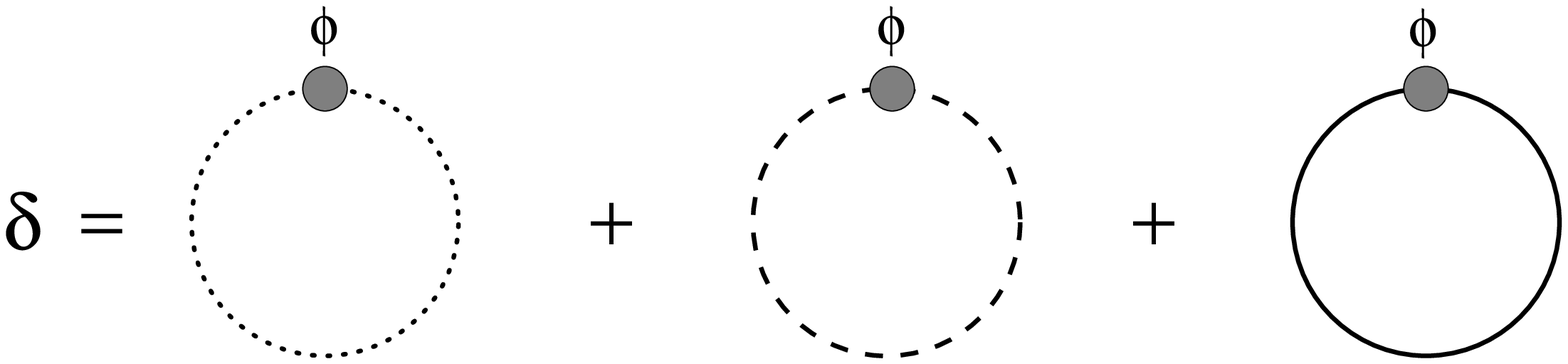}}
\caption{\it\baselineskip=12pt
The $\delta$-type amputated one-point function}
\bigskip
\label{delta}\end{figure}

Let us now briefly compare this approach
to the one of Ref. \cite{Brandem}, whose authors have used a more
intuitive method by computing the energy-momentum tensor
of the quantum fluctuations and their
back-reaction over the original backgrounds directly.
The quantum fluctuations are obtained by solving the linear order
perturbation
equations and the  constraints. In our  approach,
as in \cite{AW1}, one uses instead the  constraints and the
gauge-fixing conditions to obtain the Feynman propagators
entering the loops. In \cite{Brandem} the back-reaction was computed
  through systematic corrections of Einstein's equations,
the first non-vanishing correction being
quadratic in the Planck length $\ell_P$, the small
expansion parameter characterizing quantum fluctuations to leading
order. The counterpart to this in our approach
is that the same parameter, $\ell_P^2$,
controls the loop expansion in quantum gravity.   Incidentally,
 this last observation neatly solves an apparent paradox:
classical  gravitational
waves always carry positive energy while, as we shall see,
the back-reaction from the quantum production of gravitational
 waves from the vacuum
turns out to have, at least in some cases, negative energy.
 The point is that, in
the case of the amplification of vacuum fluctuations,
there is no parameter allowing a separation of real-graviton
production from other virtual one-loop effects whose sign
is  undetermined. This a-priori non-determination of the sign of the
back-reaction is self-evident in the effective action approach
 \cite{Vilko}.

We will follow the same procedure as outlined above in a different gauge
where physical variables are more explicit.
The usefulness of this gauge was first pointed out in \cite{Huang} and
its significance has been explored in detail in \cite{GMGG}, in the
context of linear order perturbations  to PBB backgrounds. In
the next section we summarize results in the off-diagonal gauge relevant
for the background (\ref{solution}). In the Appendix we will comment on
the gauge dependence of the results by comparing them to those
obtained in other gauges, e.g.
in the covariant gauge of \cite{Iliop}.

\newpage

\section{Propagators in the off-diagonal gauge}
\setcounter{equation}{0}

As explained in the previous section, in order to
 analyse the perturbations we need to expand (\ref{action}) to quadratic
and cubic orders in the fluctuations. We find:
\beq
&&\!\!\!\!\!\!\!\!\!
S=\int d^4x\sqrt {-g}\bigg[
{g}^{\alpha \beta} { g}^{\rho \sigma}
{ g}^{\mu \nu} \bigg\{
\frac{1}{2} \psi_{\alpha \rho , \mu} \psi_{\nu \sigma , \beta}
- \frac{1}{2} \psi_{\alpha \beta, \rho}  \psi_{\sigma \mu , \nu}
+ \frac{1}{4} \psi_{\alpha \beta , \rho} \psi_{\mu \nu , \sigma}\nn
&&\!\!\!\!\!\!
- \frac{1}{4} \psi_{\alpha \rho , \mu} \psi_{\beta \sigma , \nu}
\bigg\}\Omega^4-\frac{1}{2}
{g}^{\rho \sigma} {g}^{\mu \nu}
\psi_{\rho \sigma , \mu} \psi_{\nu}^{~\alpha}
(\Omega^2)_{, \alpha}
-{\varphi'_0} \partial_\nu \phi \, g^{0 \nu}
-\frac{1}{2} \partial_\mu \phi \, \partial_\nu \phi
\, g^{\mu \nu}\bigg] \label{fullaction}
\eeq
Here the `pseudo-graviton' ($\psi_{\mu \nu}$) indices are raised and
lowered by $\eta_{\mu \nu}$. To complete the expansion of the
$g_{\mu \nu}$ fields in terms of the background and the pseudo-graviton,
we recall the expansions
\beq
g^{\mu \nu}&=&\Omega^{-2} (\eta^{\mu \nu}-\psi^{\mu \nu}+\psi^{\mu \sigma}
\psi_{\sigma}^{\nu}+...) \nonumber \\
\sqrt{-g}&=&\Omega^4 \left[1+\frac{1}{2} \psi + (\frac{1}{8} \psi^2-
\frac{1}{4} \psi^{\alpha \beta} \psi_{\alpha \beta})+...\right]~,
\eeq
where $\psi=\psi^\mu_\mu$.

Following \cite{physrep}
we split the fluctuations into  tensor and scalar parts by
\beq
\psi_{\mu\nu}=\left(\matrix{0&0\cr 0&h_{ij}\cr}\right) + \left(\matrix{
2\Phi&-\partial_iB\cr-\partial_iB&2(\Psi\delta_{ij}-\partial_i\partial_jE)}
\right) ~, \label{fluct}
\eeq
where $h_{ij}$ represents the (transverse and traceless) tensor part with 
$h_{ii}=0$ and $\partial_ih_{ij}=0$. The trace $\psi$ is given by
$\psi=-2\Phi+6\Psi-2\nabla^2E$, where $\nabla^2\equiv
\partial_i\partial_i$. The action (\ref{action}), when expanded up to
quadratic order in the variables (\ref{fluct}), shows that
$B$ and $\Phi$ behave like Lagrange
multipliers, giving rise to the following constraints
(a prime refers to $\partial/\partial\eta$):
\beq
B&:&\quad\Psi'+{\cal H}\Phi+{1\over 4}\varphi'_0\phi=0\nn
\Phi&:&\quad 4\nabla^2\Psi-12{\cal H}\Psi'+\varphi'_0\phi'-4{\cal H}
\nabla^2(B-E')=0\;.\label{constr}
\eeq
Making use of the first constraint in (\ref{constr}), it is possible
to diagonalize the quadratic part of the action in our background as
\beq
S_2={1\over 2}\int d^4x\left[\Big(v'v'-\partial_iv\partial_iv+{\Omega''\over
\Omega}v^2\Big)+{1\over 2}\Big(\zeta'_{ij}\zeta'_{ij}-\partial_k\zeta_{ij}
\partial_k\zeta_{ij}+{\Omega''\over\Omega}\zeta_{ij}\zeta_{ij}\Big)\right]\;,
\eeq
where $v$ and $\zeta_{ij}$ are the canonical variables,
given in terms of the original fields as
\beq
v=\Omega(\phi-{\varphi'_0\over{\cal H}}\Psi)=\Omega(\phi+2\sqrt 3\Psi)\,,
\qquad\zeta_{ij}=\Omega h_{ij}\;.
\eeq
Owing to two local gauge symmetries that preserve the form
(\ref{fluct}), two out of the four
 functions can be fixed through two independent gauge
choices. The off-diagonal gauge  corresponds to choosing $\Psi=E=0$,
which simplifies the first canonical variable to
$v=\Omega\phi$. The rest of the variables become related to this
canonical variable in simple ways because of other
relations following from the constraints (\ref{constr}),  $\Phi=\sqrt
3\phi/2$ and $\nabla^2B=-\sqrt 3\phi'/2$.

The process of obtaining the propagator for the canonical variables
is simplified by
dragging the scale factor, $\Omega$, from the canonical variables to the
quadratic differential operator. In the off-diagonal
gauge this amounts to
\beq
S_2={1\over 2}\int d^4x\left[\phi\Delta\phi+{1\over
2}h_{ij}\Delta h_{ij}\right]\,,
\qquad
\Delta=\Omega\left[-\partial_\eta^2+\nabla^2+{\Omega''\over\Omega}\right]
\Omega\label{actiontwo}.
\eeq
The differential operator $\Delta$, in the zero-momentum limit, is
easily inverted:
\beq
\Delta^{-1}_0f(\eta)\equiv -\int_{\eta_i}^\eta d\eta_1\Omega^{-2}(\eta_1)
\int_{\eta_i}^{\eta_1} d\eta_2 f(\eta_2)\,,\quad
\Delta_0=\Omega\left[-\partial_\eta^2+{\Omega''\over\Omega}\right]
\Omega\label{invert},
\eeq
as has been advertised in \cite{AW2}. This choice of the inverse
operator also enforces the boundary conditions
\beq
\Delta^{-1}_0f(\eta_i)=(\Delta^{-1}_0f)'(\eta_i)=0,
\eeq
a property we require for the corrections $A$, $C$ and $D$.

From the quadratic action (\ref{actiontwo}) it is easy to read out the
Feynman propagators for the physical modes
\beq
\langle\phi(x)\phi(y)\rangle&=&i\Delta^{-1}(x,y)\nn \langle
h_{ij}(x)h_{kl}(y)\rangle&=&2T_{ijkl}\left[i\Delta^{-1}(x,y)\right]\;,
\eeq
where $T_{ijkl}$ is an appropriately normalized tensor enforcing the
transverse and traceless conditions on gravitons, i.e.
$\partial_iT_{ijkl}=\partial_jT_{ijkl}
=\partial_kT_{ijkl}
=\partial_lT_{ijkl}=0$ and $T_{iikl}=T_{ijkk}=0$.
For our purpose we are interested in computing the propagator
$\Delta^{-1}(x,y)$ in the coincident limit $x\to y$.
To do this we go to the Fourier-transformed variables
\beq
\phi(x)=\int{d^3k\over(2\pi)^3}\phi_{\bf k}(\eta)e^{i{\bf k\cdot x}}\,,
\qquad h_{ij}(x)=\int{d^3k\over(2\pi)^3}h_{{\bf k}ij}(\eta)e^{i{\bf k\cdot
x}}~~,
\eeq
where both the mode functions $\phi_{\bf k}$ and $h_{{\bf k}ij}$ can be
obtained by solving the differential equations $\Delta f_{\bf
k}(\eta)=0$. The solutions can be expressed in terms of Hankel
functions of the second
kind. Notice that, \,$v=\Omega\phi$\,
being the canonical variable, one should
normalize its modes in the far past, $\eta\to-\infty$, as $1/\sqrt|{\bf
k}|$. By the large-argument limit of the
Hankel function, this completely fixes the modes:
\beq f_{\bf
k}(\eta)={1\over 2\Omega}\sqrt{\pi\eta}H^{(1)}_0(|{\bf k}\eta|) \; .
\eeq
Consequently the propagator in the coincident-point limit takes the
 form
\beq
\lim_{x\to y}\langle\phi(x)\phi(y)\rangle&=&{|\eta|\over
8\pi\Omega^2}\int dkk^2
H^{(1)*}_0(k|\eta|)H^{(1)}_0(k|\eta|)\nn
\lim_{x\to y}\langle h_{ij}(x)h_{kl}(y)\rangle&=&{|\eta|\over
4\pi\Omega^2}\int dkk^2
(\delta_{ik}\delta_{jl}+\delta_{il}\delta_{jk}-\delta_{ij}\delta_{kl}-\delta_{ik}
{k_jk_l\over k^2}-\delta_{jl}{k_ik_k\over k^2}\nn
&&-\delta_{il}{k_jk_k\over k^2}-
\delta_{jk}{k_ik_l\over k^2}+\delta_{kl}{k_ik_j\over k^2}+\delta_{ij}
{k_kk_l\over k^2}+{k_ik_jk_kk_l\over k^4})\nn &&\times
H^{(1)*}_0(k|\eta|)H^{(1)}_0(k|\eta|)\;.\label{prop}
\eeq
Here we have replaced the transverse, traceless $T_{i j k l}$ tensor
by its Fourier transform. %\cite{Ford}.
The integral in (\ref{prop}) is in general divergent in the
ultraviolet. It should be kept in mind, however,
that the present approach should be complemented by
a momentum cutoff as a result
of the ultraviolet sickness of gravity.
We set the cutoff, at any given time, at
$k_{max} \sim \Omega |H|$, which corresponds
to a wavelength of the order of the  Hubble radius. This is
also the scale at which the mode functions switch from
free plane waves to `frozen out' modes.
For modes larger than this wavelength the bosonic contribution dominates 
its fermionic counterpart (we assume some underlying
supersymmetry) as a
result of Bose-condensation. This is the effect we are after.
For shorter-wavelength modes new degrees of freedom
should appear, which would ultimately
provide better ultraviolet behaviour, as in superstring theory.
 We also impose an infrared cutoff coming from a
finite duration of dilaton-driven inflation, $k_{min} \sim |H_i|$,
corresponding to the maximal Hubble radius.
In conclusion  we restrict  momenta in all loops to be
in the range: $|H_i|\le k\le \Omega |H|=|H_i|/\Omega^2$. At this
upper-momentum/short-wavelength cutoff, $\Omega |H| \sim
1/|\eta|$, i.e. our cutoff corresponds to $|k \eta| \sim 1$.
Note that, in the off-diagonal gauge, as a result of the constraints, 
all fields can be expressed in terms of $\phi$. Thus only
one type of amputated loops, the $\delta$-type, effectively arises 
on the right-hand side of (\ref{final}).

\section{Computation of the one-point functions}
\setcounter{equation}{0}

In the first two subsections we shall compute the amputated one-point  
functions
due to graviton, dilaton, and gauge-boson loops. In the last subsection
the external zero-momentum propagators will be glued in to reconstruct the
full one-point functions.

\subsection{Graviton and dilaton loops}

To get the three-point vertices we use our expansion of the action
(\ref{fullaction}) to
cubic order in the fluctuations, $S_3$. The results are summarized
in tables 1 and 2, showing all the relevant three-point vertices.
In table 3 we list all the
coincident loops appearing in $\delta$ and their respective
contributions to the loop integrals ($x=|k\eta|$). There is another  
way of doing the same computation, which involves, in
general, four types of amputated loops in the off-diagonal gauge (Fig. 5).
\begin{figure}[htb]
\epsfxsize=3.5in
\bigskip
\centerline{\epsffile{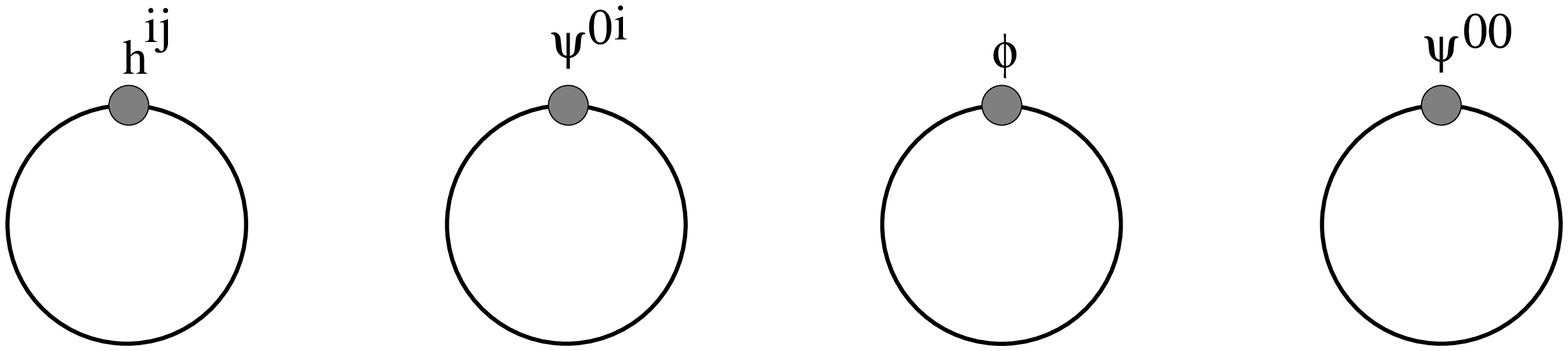}}
\caption{\it\baselineskip=12pt
All possible types of amputated one-point functions in the
off-diagonal gauge.
Here the line represents all possible coincident propagators.}
\bigskip
\label{offd}\end{figure}

The first type of loops, having $h_{ij}$ in the external leg, turn
out to be
zero, since the loop (see both tables 1 and 2),
in the coincident limit,
always picks up a factor of $\delta_{ij}$, which
gives vanishing contributions  because of the tracelessness of
$h_{ij}$. The second type of loops, having $\Psi_{0i}$ in the external
leg, also gives vanishing contributions, but for a different
reason. Note that $\Psi_{0i}=\partial_iB$ can be integrated
by parts to act on the loop. The loops, however, being functions only of
time in the coincident limit (\ref{prop}) give zero when hit by a space
derivative. Thus only the third and fourth types of loops survive in the
off-diagonal gauge, the $\delta$-type and $\gamma$-type loops as
defined in Figs. 3 and 4.

\vspace{1cm}
%\newpage
\begin{tabular}{|l|l|r|}\hline

1&
${1\over 2}H_i\partial_2^{\alpha_3}\eta^{\alpha_1
\beta_1}\eta^{
 \alpha_2
 \beta_2}\psi_{
 \alpha_1
 \beta_1}(1)\psi_{
 \alpha_2
 \beta_2}(2)\psi_{
 \alpha_3
 \beta_3}(3)t^
 {\beta_3}$\\ \hline

2&
 $-H_i\partial_2^
 {
 \alpha_3}\eta^
 {
 \alpha_1
 \alpha_2}\eta^
 {
 \beta_2
 \beta_1}\psi_{
 \alpha_1
 \beta_1}(1)\psi_{
 \alpha_2
 \beta_2}(2)\psi_{
 \alpha_3
 \beta_3}(3)t^
 {
 \beta_3}$\\ \hline

3&
 $-H_i\partial_2^
 {
 \beta_1}\eta^
 {
 \alpha_2
 \beta_2}\eta^
 {
 \beta_3
 \alpha_1}\psi_{
 \alpha_1
 \beta_1}(1)\psi_{
 \alpha_2
 \beta_2}(2)\psi_{
 \alpha_3
 \beta_3}(3)t^
 {
 \alpha_3}$\\ \hline

4&
 ${1\over 4}\Omega^2\partial_2^
 {
 \beta_3}\partial_3^
 {
 \alpha_2}\eta^
 {
 \alpha_1
 \beta_1}\eta^
 {
 \beta_2
 \alpha_3}\psi_{
 \alpha_1
 \beta_1}(1)\psi_{
 \alpha_2
 \beta_2}(2)\psi_{
 \alpha_3
 \beta_3}(3)$\\ \hline

5&
 $-
 \Omega^2\partial_2^
 {
 \beta_3}\partial_3^
 {
 \alpha_1}\eta^
 {
 \beta_1
 \alpha_2}\eta^
 {
 \beta_2
 \alpha_3}\psi_{
 \alpha_1
 \beta_1}(1)\psi_{
 \alpha_2
 \beta_2}(2)\psi_{
 \alpha_3
 \beta_3}(3)$\\ \hline

6&
 ${1\over 2}
 \Omega^2\partial_2^
 {
 \beta_3}\partial_3^
 {
 \alpha_2}\eta^
 {
 \beta_1
 \alpha_3}\eta^
 {
 \beta_2
 \alpha_1}\psi_{
 \alpha_1
 \beta_1}(1)\psi_{
 \alpha_2
 \beta_2}(2)\psi_{
 \alpha_3
 \beta_3}(3)$\\ \hline

7&
 ${1\over 4}
 \Omega^2\partial_2^
 {
 \alpha_3}\partial_3^
 {
 \beta_3}\eta^
 {
 \alpha_1
 \beta_1}\eta^
 {
 \alpha_2
 \beta_2}\psi_{
 \alpha_1
 \beta_1}(1)\psi_{
 \alpha_2
 \beta_2}(2)\psi_{
 \alpha_3
 \beta_3}(3)$\\ \hline

8&
 ${1\over 2}\Omega^2\partial_2^
 {
 \alpha_3}\partial_3^
 {
 \beta_3}\eta^
 {
 \alpha_1
 \alpha_2}\eta^
 {
 \beta_2
 \beta_1}\psi_{
 \alpha_1
 \beta_1}(1)\psi_{
 \alpha_2
 \beta_2}(2)\psi_{
 \alpha_3
 \beta_3}(3)$\\ \hline

9&
 ${1\over 2}\Omega^2\partial_2^
 {
 \alpha_1}\partial_3^
 {
 \beta_3}\eta^
 {
 \alpha_2
 \beta_2}\eta^
 {
 \beta_1
 \alpha_3}\psi_{
 \alpha_1
 \beta_1}(1)\psi_{
 \alpha_2
 \beta_2}(2)\psi_{
 \alpha_3
 \beta_3}(3)$\\ \hline

10&
 ${1\over 2}\Omega^2\partial_2^
 {
 \alpha_1}\partial_3^
 {
 \beta_2}\eta^
 {
 \alpha_3
 \beta_3}\eta^
 {
 \beta_1
 \alpha_2}\psi_{
 \alpha_1
 \beta_1}(1)\psi_{
 \alpha_2
 \beta_2}(2)\psi_{
 \alpha_3
 \beta_3}(3)$\\ \hline

11&
 ${1\over 8}\Omega^2\partial_2^
 {
 \alpha_4}\partial_3^
 {
 \alpha_5}\eta_{
 \alpha_4
 \alpha_5}\eta^
 {
 \alpha_1
 \beta_1}\eta^
 {
 \alpha_2
 \beta_2}\eta^
 {
 \alpha_3
 \beta_3}\psi_{
 \alpha_1
 \beta_1}(1)\psi_{
 \alpha_2
 \beta_2}(2)\psi_{
 \alpha_3
 \beta_3}(3)$\\ \hline

12&
 ${1\over 2}
 \Omega^2\partial_2^
 {
 \alpha_4}\partial_3^
 {
 \alpha_5}\eta_{
 \alpha_4
 \alpha_5}\eta^
 {
 \alpha_1
 \alpha_2}\eta^
 {
 \alpha_3
 \beta_3}\eta^
 {
 \beta_2
 \beta_1}\psi_{
 \alpha_1
 \beta_1}(1)\psi_{
 \alpha_2
 \beta_2}(2)\psi_{
 \alpha_3
 \beta_3}(3)$\\ \hline

13&
 ${1\over 4}
 \Omega^2\partial_2^
 {
 \alpha_1}\partial_3^
 {
 \beta_1}\eta^
 {
 \alpha_2
 \beta_2}\eta^
 {
 \alpha_3
 \beta_3}\psi_{
 \alpha_1
 \beta_1}(1)\psi_{
 \alpha_2
 \beta_2}(2)\psi_{
 \alpha_3
 \beta_3}(3)$\\ \hline

14&
 ${1\over 8}
 \Omega^2\partial_2^
 {
 \alpha_4}\partial_3^
 {
 \alpha_5}\eta_{
 \alpha_4
 \alpha_5}\eta^
 {
 \alpha_1
 \beta_1}\eta^
 {
 \alpha_2
 \alpha_3}\eta^
 {
 \beta_3
 \beta_2}\psi_{
 \alpha_1
 \beta_1}(1)\psi_{
 \alpha_2
 \beta_2}(2)\psi_{
 \alpha_3
 \beta_3}(3)
 $\\ \hline

15&
 ${1\over 2}\Omega^2\partial_2^
 {
 \alpha_4}\partial_3^
 {
 \alpha_5}\eta_{
 \alpha_4
 \alpha_5}\eta^
 {
 \alpha_1
 \alpha_2}\eta^
 {
 \beta_2
 \alpha_3}\eta^
 {
 \beta_3
 \beta_1}\psi_{
 \alpha_1
 \beta_1}(1)\psi_{
 \alpha_2
 \beta_2}(2)\psi_{
 \alpha_3
 \beta_3}(3)$\\ \hline

16&
 ${1\over 4}\Omega^2\partial_2^
 {
 \alpha_1}\partial_3^
 {
 \beta_1}\eta^
 {
 \alpha_2
 \alpha_3}\eta^
 {
 \beta_3
 \beta_2}\psi_{
 \alpha_1
 \beta_1}(1)\psi_{
 \alpha_2
 \beta_2}(2)\psi_{
 \alpha_3
 \beta_3}(3)$\\ \hline

\end{tabular}\\[.5cm]
Table 1: {\em List of all $\psi_{\alpha_1\beta_1}\psi_{\alpha_2\beta_2}
\psi_{\alpha_3\beta_3}$ vertices. While choosing an external leg one
should permute $1,2,3$ in each vertex.}

\vspace{1cm}

\begin{tabular}{|l|l|r|}\hline

1&
$-{\sqrt 3\over 4}
 H_i\partial_1^
 {
 \alpha_6}\eta^
 {
 \alpha_4
 \alpha_5}\eta^
 {
 \beta_4
 \beta_5}\phi(1)\psi_{
 \alpha_4
 \alpha_5}(2)\psi_{
 \beta_4
 \beta_5}(3)t_{
 \alpha_6}
 $\\ \hline

2&
 ${\sqrt 3\over 2}H_i\partial_1^
 {
 \alpha_4}\eta^
 {
 \alpha_5
 \beta_5}\eta^
 {
 \alpha_6
 \beta_6}\phi(1)\psi_{
 \alpha_5
 \alpha_6}(2)\psi_{
 \beta_5
 \beta_6}(3)t_{
 \alpha_4}$\\ \hline

3&
 $\sqrt 3H_i\partial_1^
 {
 \alpha_5}\eta^
 {
 \alpha_7
 \alpha_8}\phi(1)\psi_{
 \alpha_4
 \alpha_5}(2)\psi_{
 \alpha_7
 \alpha_8}(3)t^
 {
 \alpha_4}$\\ \hline

4&
 $-2\sqrt 3H_i\partial_1^
 {
 \alpha_5}\eta^
 {
 \alpha_6
 \alpha_7}\phi(1)\psi_{
 \alpha_4
 \alpha_6}(2)\psi_{
 \alpha_5
 \alpha_7}(3)t^
 {
 \alpha_4}$\\ \hline

5&
 $-{1\over 4}
 \Omega^2\partial_1^
 {
 \alpha_4}\partial_2^
 {
 \alpha_5}\eta_{
 \alpha_4
 \alpha_5}\eta^
 {
 \alpha_6
 \alpha_7}\phi(1)\phi(2)\psi_{
 \alpha_6
 \alpha_7}(3)
 $\\ \hline

6&
 ${1\over 2}\Omega^2\partial_1^
 {
 \alpha_4}\partial_2^
 {
 \alpha_5}\phi(1)\phi(2)\psi_{
 \alpha_4
 \alpha_5}(3)$\\ \hline

\end{tabular}\\[.5cm]
Table 2: {\em All possible $\phi\psi_{\alpha_1\beta_1}
\psi_{\alpha_2\beta_2}$ and $\phi\phi\psi_{\alpha_1\beta_1}$ vertices.
Again one should permute appropriate indices when considering each
vertex.}

\vspace{1cm}

\begin{tabular}{|l|l|r|}\hline

$\langle\phi,\phi'\rangle$ &
$\int dxx^3H^{(1)*}_0(x)H'^{(1)}_0(x)/16H_i\pi|\eta|^4$\\ \hline

$\langle\phi,\phi''\rangle$ &
$-\int dxx^4H^{(1)*}_0(x)H''^{(1)}_0(x)/16H_i\pi|\eta|^5$\\ \hline

$\langle\partial_i\phi,\partial_i\phi\rangle$ &
$-\int dxx^4H^{(1)*}_0(x)H^{(1)}_0(x)/16H_i\pi|\eta|^5$\\ \hline

$\langle\partial_i\phi,\psi_{0i}\rangle$ &
$-\sqrt 3\int dxx^3H^{(1)*}_0(x)H'^{(1)}_0(x)/32H_i\pi|\eta|^4$\\ \hline

$\langle\partial_i\phi,\psi'_{0i}\rangle$ &
$\sqrt 3\int dxx^4H^{(1)*}_0(x)H''^{(1)}_0(x)/32H_i\pi|\eta|^5$\\ \hline

$\langle\partial_i\phi',\psi_{0i}\rangle$ &
$\sqrt 3\int dxx^4H'^{(1)*}_0(x)H'^{(1)}_0(x)/32H_i\pi|\eta|^5$\\ \hline

$\langle\psi_{0i},\psi'_{0i}\rangle$ &
$-3\int dxx^3H'^{(1)*}_0(x)H''^{(1)}_0(x)/64H_i\pi|\eta|^4$\\ \hline

$\langle\partial_i\psi_{0j},\partial_j\psi_{0i}\rangle$ &
$-3\int dxx^4H'^{(1)*}_0(x)H'^{(1)}_0(x)/64H_i\pi|\eta|^5$\\ \hline

$\langle\partial_i\psi_{0j},\partial_i\psi_{0j}\rangle$ &
$-3\int dxx^4H'^{(1)*}_0(x)H'^{(1)}_0(x)/64H_i\pi|\eta|^5$\\ \hline

$\langle\psi_{ij},\psi'_{ij}\rangle$ &
$3\int dxx^3H^{(1)*}_0(x)H'^{(1)}_0(x)/8H_i\pi|\eta|^4$\\ \hline

$\langle\partial_i\psi_{jk},\partial_j\psi_{ik}\rangle$ &
$0$\\ \hline

$\langle\partial_i\psi_{jk},\partial_i\psi_{jk}\rangle$ &
$-3\int dxx^4H^{(1)*}_0(x)H^{(1)}_0(x)/8H_i\pi|\eta|^5$\\ \hline

$\langle\phi',\phi'\rangle$ &
$-\int dxx^4H'^{(1)*}_0(x)H'^{(1)}_0(x)/16H_i\pi|\eta|^5$\\ \hline

$\langle\psi'_{ij},\psi'_{ij}\rangle$ &
$-3\int dxx^4H'^{(1)*}_0(x)H'^{(1)}_0(x)/8H_i\pi|\eta|^5$\\ \hline

\end{tabular}\\[.5cm]
Table 3: {\em All possible coincident loop propagators appearing for
the off-diagonal gauge. The prime in $H'^{(2)}_0$ refers to $\partial
/\partial x$, otherwise to $\partial/\partial\eta$.}
\vspace{1cm}

Another important feature of the off-diagonal gauge is the absence of the
Faddeev--Popov ghosts. This can be readily seen from the gauge
transformations of the gauge-fixing functions $\Psi$ and $E$.
Their gauge transformations have been given in \cite{physrep}:
\beq
\Psi\to\Psi+{\cal H}\xi(x)\,,\qquad E\to E-\overline\xi(x)\;,
\eeq
where $\xi$ and $\overline\xi$ are the two gauge parameters associated
with the coordinate transformations: $\eta\to\eta-\xi$ and $x^i\to x^i-
\partial_i\overline\xi$. Clearly, the Faddeev--Popov determinant is
field-independent and hence the ghosts are decoupled, unlike what happens in
the covariant gauge of \cite{Iliop}.

As an example, we present a sample calculation, say the three-point vertex
$\#$5 of table 2. For illustrative purposes, it is easier to make
use of the
constraints at the beginning (in contrast, in the computer code
its more convenient to impose the constraints later). As we have argued,  
this does not make
a difference, which is at the same time an important consistency check  
for our method. Making use of the constraints in the off-diagonal gauge  
presented in
the previous section, the ${\rm tr}(\psi_{\mu\nu})$, present in the vertex
$\#$5, can be replaced by the scalar fluctuation $\phi$. As we said, in
this case there is only one type of amputated loop $\delta_{\rm
eff}$, that
can be read off from the vertex $\#$5:
\beq
\delta_{\rm eff}={\sqrt 3\over 4}[-\Omega^2\langle\phi'\phi'\rangle+4H_i
\langle\phi\phi'\rangle+\Omega^2\langle\phi\phi\rangle''+\Omega^2\langle
\partial_i\phi\partial_i\phi\rangle]\;.\label{example}
\eeq
Now, one should look at table 3, where various propagators have been
listed in this gauge and pick up the leading-order terms from
(\ref{example})
as we approach the limit,
$\eta\to 0^-$. Recall that $\Omega^2\sim|\eta|$, and hence all the terms
in (\ref{example}) are of the same order,  $1/\eta^4$.

\subsection{Gauge-boson loops}

Let us now study the back-reaction
produced by gauge fields to the background
(\ref{solution}). There are
two distinct types of
gauge fields appearing in the low-energy string effective action ---
the NS-type and the R-type. The NS-type couple to
 all the fundamental moduli while the R-type
couple only to  the metric. Let us first
concentrate on the NS-type gauge bosons. In the
canonical frame the action is:
\beq
S^{\rm em}={1\over 4}\int d^4xe^{-\varphi}\sqrt{-g}g^{\mu\rho}g^{\nu\sigma}
F_{\mu\nu}F_{\rho\sigma}\;,\label{emaction}
\eeq
which supplements the graviton--dilaton action (\ref{action}) in the  
presence
of the NS-type fields. Since the gauge-boson action
(\ref{emaction}) is already quadratic in the gauge fields,
these decouple,
to linear order, from other fluctuations. Also, interactions stemming from
a possible non-Abelian gauge structure can be ignored.
We shall work in the  Coulomb gauge where the temporal component
of the gauge field is zero and the spatial components  satisfy
 $\partial_iA_i=0$. Going over to a
canonical gauge-field variable, the action can be written in the form
\beq
S^{\rm em}_2={1\over 2}\int d^4x\zeta_i\overline\Delta\zeta_i\,,\qquad
\overline\Delta=-\partial_\eta^2+\nabla^2-e^{\varphi_0/2}(e^{-
\varphi_0/2})''\;,
\eeq
where the canonical field is given by $\zeta_i=e^{-\varphi_0/2}A_i$.
Fourier expanding as before, the  mode
functions satisfy the homogeneous equation
\beq
\zeta_i=\int{d^3k\over(2\pi)^3}\zeta_{{\bf k}i}(\eta)e^{\bf ik\cdot x}\,,
\quad\overline\Delta\zeta_i=0\,,\quad{\bf k}^i\zeta_{{\bf k}i}=0\;,
\eeq
the solution of which can be written down \cite{gvv} in terms of the Hankel
functions of the second kind. Following the same normalization arguments
as presented before, the Coulomb-gauge Feynman propagator
for the canonical fields
 can be expressed in the form
\beq
\!\!\!\!\langle\zeta_i(x)\zeta_j(y)\rangle=\int{d^3k\over(2\pi)^3}{1
\over 4}
\Big(\delta_{ij}-{k^ik^j\over  
k^2}\Big)\left[\sqrt{\pi\eta}H^{(1)}_\lambda(kx^0)
\right]\left[\sqrt{\pi\eta}H^{(1)*}_\lambda(ky^0)\right]e^{i{\bf  
k\cdot (x-y)}} \eeq
where the index $\lambda=(\sqrt 3-1)/2$. The cubic vertices have been listed
in table 4 and the coincident loops are presented in table 5.

For the R-type gauge bosons, the coupling to the dilaton being absent  
in the effective action
(\ref{emaction}), the mode functions satisfy the free wave equations:
$\zeta_{{\bf k}i}''+k^2\zeta_{{\bf k}i}=0$. Clearly, these modes are not
parametrically amplified; and hence their contributions to the  
back-reaction  are negligible.

The relevant diagrams for gauge bosons are the same as in Fig.  
\ref{offd} except
that here the solid line represents the gauge loop. Notice, from table 4,
that all the three-point vertices are quadratic in the gauge fields  
and at the
linear order there is no coupling between the gauge fields and the
gravitational
perturbations. Among the loops, see again Fig. \ref{offd}, the ones  
that have
$h_{ij}$ and $\psi_{0i}$ as external legs are zero for the same reason as
presented before. The rest of
the diagrams are of the $\gamma$ or $\delta$ type. As an example, let us
take the vertex $\#$6 of table 4. It contributes to $\delta_{\rm eff}$
as
\beq
\delta_{\rm eff}={1\over 2}\exp(-\varphi_0)[-\langle A'_iA'_i\rangle+
\langle\partial_kA_i\partial_kA_i\rangle]\;.
\eeq
The exponential factor is common to all the vertices and is
$\sim|\eta|^{\sqrt 3}$, which cancels the inverse factor in the second
coincident-limit propagator, as can be seen from table 5. It requires some
more effort to see that this is also happening for the first entry
 of table 5, but
 again, even for gauge bosons, the leading effect
in the amputated one-point functions is $\sim 1/\eta^4$.

\newpage
%\vspace{1cm}

\begin{tabular}{|l|l|r|}\hline

1&
 $-{1\over 4}\exp(-\varphi_0)\partial_{2}^
 {
 \alpha_5}\partial_{3}^
 {
 \beta_6}\eta_{
 \alpha_5
 \beta_6}\eta^
 {
 \alpha_4
 \beta_4}\eta^
 {
 \beta_5
 \alpha_6}A_{
 \beta_5}(2)A_{
 \alpha_6}(3)\psi_{
 \alpha_4
 \beta_4}(1)
 $ \\ \hline

2&
 ${1\over 4}\exp(-\varphi_0)\partial_{2}^
 {
 \alpha_5}\partial_{3}^
 {
 \beta_5}\eta^
 {
 \alpha_4
 \beta_4}A_{
 \beta_5}(2)A_{
 \alpha_5}(3)\psi_{
 \alpha_4
 \beta_4}(1)$ \\ \hline

3&
 ${1\over 2}\exp(-\varphi_0)\partial_{2}^
 {
 \alpha_5}\partial_{3}^
 {
 \beta_5}\eta^
 {
 \alpha_6
 \beta_6}A_{
 \alpha_6}(2)A_{
 \beta_6}(3)\psi_{
 \alpha_5
 \beta_5}(1)$ \\ \hline

4&
 ${1\over 2}\exp(-\varphi_0)\partial_{2}^
 {
 \beta_5}\partial_{3}^
 {
 \alpha_5}\eta_{
 \alpha_5
 \beta_5}\eta^
 {
 \alpha_4
 \alpha_6}\eta^
 {
 \beta_4
 \beta_6}A_{
 \alpha_6}(2)A_{
 \beta_6}(3)\psi_{
 \alpha_4
 \beta_4}(1)$ \\ \hline

5&
 $-\exp(-\varphi_0)\partial_{2}^
 {
 \alpha_5}\partial_{3}^
 {
 \beta_4}\eta^
 {
 \alpha_4
 \beta_5}A_{
 \beta_5}(2)A_{
 \alpha_5}(3)\psi_{
 \alpha_4
 \beta_4}(1)
 $ \\ \hline

6&
 ${1\over 2}\exp(-\varphi_0)\partial_2^
 {
 \alpha_5}\partial_{3}^
 {
 \beta_6}\eta_{
 \alpha_5
 \beta_6}\eta^
 {
 \beta_5
 \alpha_6}\phi(1)A_{
 \beta_5}(2)A_{
 \alpha_6}(3)$  \\ \hline

7&
 $-{1\over 2}
 \exp(-\varphi_0)\partial_2^{
 \alpha_5}\partial_3^
 {
 \beta_5}\phi(1)A_{
 \beta_5}(2)A_{
 \alpha_5}(3)
 $\\ \hline

\end{tabular}\\[.5cm]
Table 4: {\em All possible $\phi A_{\alpha_1}
A_{\beta_1}$ and $\psi_{\alpha_1\beta_1}A_{\alpha_2}A_{\beta_2}$ vertices.
Again one should permute indices when considering each vertex.}

\vspace{1cm}

\begin{tabular}{|l|l|r|}\hline

 $\langle
 A'_{
 i_1},A'_{
 i_1}\rangle$& $
 \int dxx^2 [\sqrt{-\eta} (\eta/\eta_i)^{-\sqrt{3}/2}
H^{(1)*}_{\lambda}(|k\eta|)]'\;\cdot$\\ &$
         [\sqrt{-\eta} (\eta/\eta_i)^{-\sqrt{3}/2}H^{(1)}_{\lambda}
(|k\eta|)]'
         /8 \pi|\eta|^3$\\ \hline

 $\langle
 \partial_
 {
 i_1}A_{
 i_2},\partial_
 {
 i_1}A_{
 i_2}\rangle$ & $
 \int dxx^4 (\eta/\eta_i)^{-\sqrt{3}} H^{(1)*}_{\lambda}(x)
H^{(1)}_{\lambda} (x)/
  8 \pi|\eta^4|
 $ \\ \hline

\end{tabular}\\[.5cm]
Table 5: {\em All coincident photon loops. All
primes refer to $\partial/\partial\eta$.}

\subsection{Adding external propagators}

Following (\ref{final}) we have first computed the amputated
one-point functions in the off-diagonal gauge. As it turned out, after
summing over all the vertices in tables 1 and 2, the leading-order
 terms, as $\eta\to 0$,
have the time dependence $1/\eta^4$. We now attach the
external line,  i.e. the inverse of the
zero-momentum kinetic operator,
$\Delta_0$. From (\ref{invert}) we note
that our amputated quantities have to go
through a double $\eta$-integral and multiplication by $\Omega^{-2}$. Thus 
  the  leading-order time
dependence of the one-point functions is simply $1/\eta^3$.
Going through the procedure more carefully, we see that,
apart for some numerical factors (which are different for different
one-point functions), the leading-order contribution to all
one-point functions is of order $1/H_i\eta^3$.
This result suggests that, in the context of PBB cosmology,  the
dominant modes are those
 near the upper momentum cut-off $k \sim \eta^{-1}$. As we have
already argued,
contributions from modes beyond this cut-off, $|k\eta|\sim 1$, have to be
combined with those from other fields, e.g. from
fermions. The resulting combined effect
will be typical of a short-distance renormalization
of the bare parameters of the Lagrangian, which should be
mild in a supersymmetric theory. Apart from this, the physical effect from
short distances should be parametrically smaller than the effect
we have computed.
Considering (\ref{solution}), (\ref{zerohubble}), and the formula
$H_i=1/2\eta_i$, the contribution we have found can be
 simply expressed, in our units, as:
\beq
\left(A,\,C,\,D\right)\sim{1\over
H_i\eta^3}\sim{H_i^2\over\Omega^6}
\equiv H_0^2\;.\label{goes}
\eeq

 What about the ratio between the three one-point functions
in the off-diagonal gauge? Our answer turns out to be simply:
\beq
A=0\,,\qquad C=\sqrt 3D\,,\qquad D=-\Delta^{-1}_0\delta_{\rm eff}\;.
\label{one-point}
\eeq
 We should
point out that, while (\ref{one-point}) appears to parametrize only certain
forms of metric fluctuations, other contributions,
such as $\langle\psi_{0i}\rangle$ or $\langle\psi_{ij} \rangle,\;i\neq j$, 
have been shown to be exactly zero. Also,
the vanishing of $\alpha$ easily follows from
the property $h_{ii}=0$ and from the coincident-point limit of the loop
(\ref{prop}).

To reach (\ref{one-point}) one could follow another path, which parallels
more closely  \cite{AW1} and \cite{AW2}.
One can take into account all types of loops --corresponding to
all the variables appearing in the fluctuations
(\ref{fluc})-- and use the constraints at the end. As a matter of
fact, this also provides a good consistency check of the method. Here
again, from very general considerations (cf. the three-point vertices in
tables 1, 2 and 4), loops with $\psi_{0i}$ and $h_{ij}$ as external legs
vanish, and only $\delta$-type and the $\gamma$-type amputated functions
survive. Thus the results (\ref{one-point}) follow in a different way:
\beq
A=0\,,\qquad C=\sqrt 3D\,,\qquad D=-\Delta^{-1}_0\left(\delta+\sqrt 3
\gamma \right)~~,
\label{one-point1}\eeq
where (\ref{one-point1}) is identical to (\ref{one-point}) since
the constraints provide the relation
 $\delta_{\rm
eff}=\delta+\sqrt 3\gamma$.

We are now ready to discuss the physical
 implications of our results for the PBB scenario.

\section{Discussion}
\setcounter{equation}{0}

The physical effects due to back-reaction are best
analysed, in a PBB context, by going over to the string frame.
In order to prepare for that,
let us write down the final outcome of our calculation
 in the off-diagonal gauge as
\beq
\langle ds^2\rangle&=&\Omega^2[-(1-C)d\eta^2+d\vec x\cdot d\vec x\,]
=-dt^2+a^2(t)d\vec x\cdot d\vec x\nn
\langle\varphi\rangle&=&\varphi_0+D\;,
\eeq
where $C$ and $D$ are as given explicitly in (\ref{one-point1}).
Since we limit ourselves to
times for which $C \ll 1$, the new comoving time $t$ is related to
the background comoving time by
\beq
dt\simeq\Omega\left(1-{1\over 2}C\right)d\eta=\left(1-{1\over  
2}C\right)dt_0\;.
\eeq
Expressing the dilaton and the Hubble parameter $H$
associated with the scale factor $a$ in the new comoving time, and after
using the relation (\ref{one-point}) between $C$ and $D$ we find
\beq
\langle\varphi\rangle&\simeq&-{2\over\sqrt 3}\left[\ln(-t)-
\sqrt 3D\right]\nn
H&=&{d\over dt}\langle\ln a\rangle\simeq{1\over 3t}\left[1+
\sqrt 3D\right]\;.\label{quadratic}
\eeq
In order to go over to the string frame, we introduce
the corresponding comoving time 
$dt_s=\exp(\langle\varphi\rangle/2)dt$ and  the
scale factor $a_s=\exp(\langle\varphi\rangle/2)a$. After a
straightforward calculation
the resulting string-frame
Hubble parameter and dilaton can be written as
\beq
H_s&=&{d\over dt_s} \langle \ln a_s \rangle \simeq-{1\over\sqrt
3t_s}\left[1+
{{7\sqrt 3}-3\over 2}D\right]\;, \nn
\langle\varphi\rangle &\simeq& -(\sqrt 3+1)\left[\ln(-t_s)-(2\sqrt 3-3)D
\right]    \;.\label{stringeffect}
\eeq
It turns out that, in the Einstein frame, $D(t)$ is a
positive number times the square of the  Hubble parameter.
 For the minimal
gravi-dilaton system, we find:
\beq
D(t) = 0.0047H_0^2(t)\;.\label{result}
\eeq
Although the coefficient of $H_0^2$ looks small, we have to recall
that, in our units ($16 \pi G = \hbar = c =1$), Planck's time is given by
$t_P^2 = 1/16\pi$. In other words, the (relative, dimensionless)
 correction to the tree-level
background is:
\beq
D=0.24\, t_P^2 H_0^2\;,\label{witht_P}
\eeq
i.e. does not contain any particularly small number.

When one goes to the string frame, in which the string length $\ell_s$
is a constant, the same result as (\ref{witht_P}) for $D$
follows, provided one interprets $t_P$ as the effective, time-dependent, 
Planck time
$t_P = \ell_s e^{\phi/2}$ and $H_0$ is expressed in terms of the
string-frame Hubble
parameter and of $\dot{\phi}$ (now the dot
refers to derivative w.r.t. cosmic string time) by the standard relation
$H_0 = H_s -\dot{\phi}/2$.

When the NS-gauge-boson loops are included the numbers in
(\ref{result}) change but
the conclusions do not. The  contribution of a single gauge boson to  
$D$ turns out to be
\beq
D=0.00016H_0^2 = 0.008~t_P^2 H_0^2\;,
\label{photonD}
\eeq
i.e. of the same sign but about $30$ times smaller
 than that of the gravi-dilaton system (\ref{result}).
The result (\ref{photonD}) should be multiplied by the number of NS
gauge bosons,
which, in typical superstring theories, could be in the range of hundreds.

In both cases the one-loop correction adds to the rate of
inflation and to the rate of
 growth of the dilaton. At  first sight, this seems to go
against an eventual exit from inflation.
 However, what is crucial for the exit is not so much the {\it absolute}
rates of growth but rather the {\it relative} rate of dilaton and
scale-factor growth.

\begin{figure}[htb]
\epsfxsize=5in
\bigskip
\centerline{\epsffile{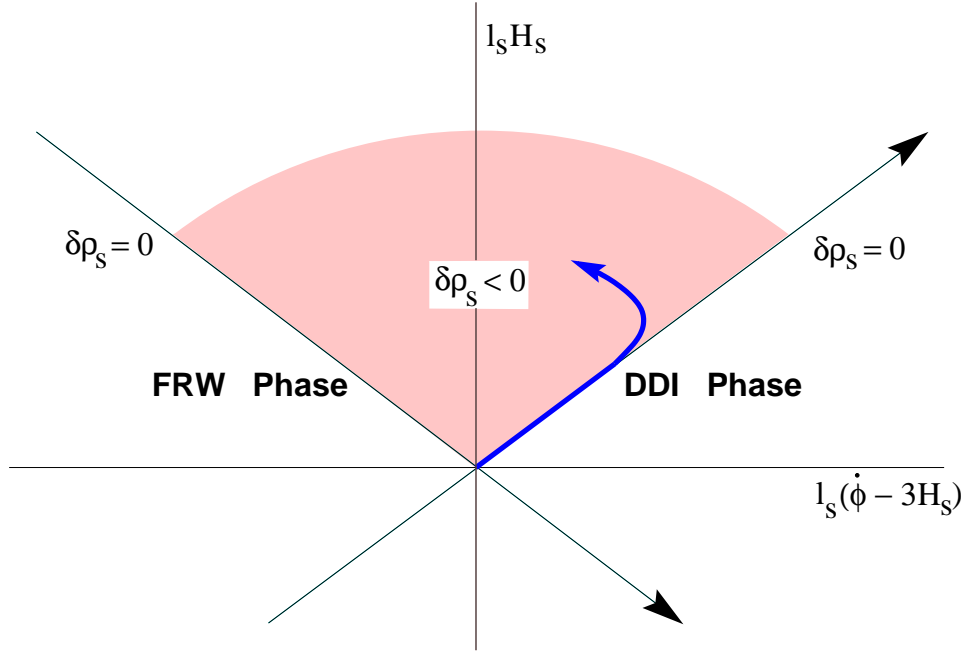}}
\caption{\it\baselineskip=12pt
The phase diagram of PBB cosmology. The curve shows the direction in  
which the back-reaction we computed is modifying the lowest order  
solution (straight line) helping an eventual flow from DDI to a  
FRW-phase.}
\bigskip
\label{exit}
\end{figure}

 To be more precise, let us compute directly
 the effect of back-reaction on the string-frame quantity
$\dot{\overline\varphi}$, where $\overline\varphi=\varphi-3\ln a_s$
is the duality-invariant shifted dilaton and, as before, the dot refers
to the time derivative in the string frame. From Eq. (\ref{stringeffect})  
we easily compute:
\beq
\dot{\overline\varphi}=-{1\over t_s}\left[1-{9-3\sqrt 3\over 2}D\right]\;.
\label{effect}\eeq
The positive sign of $D$ is forcing $\dot{\overline\varphi}$ to decrease
(relative to the tree-level value) as the result of a competition between
 a growing dilaton and a growing Hubble parameter in the string frame.
It is thus obvious that the ratio $\dot{\overline\varphi}/H_s$ will
change from its tree-level value of $\sqrt3$ towards smaller values.
As shown in  Fig. \ref{exit}, this is precisely as required by a graceful 
exit \cite{exit}, i.e.
the phase-space trajectory in the
$\dot{\overline\varphi}$-$H$ plane should turn
anticlockwise. Incidentally, this is also what the Hubble entropy bound
of \cite{entropy} requires \cite{support}.

Another way the effect (\ref{effect}) can be understood is by computing the
effective energy and pressure densities produced
by the quantum fluctuations and by checking whether
the condition $\delta \rho_s+\delta p_s < 0$ is obeyed, where
$\delta\rho_s$ and $\delta p_s$ are the
energy  and pressure densities in the quantum fluctuations.
Let us study this first in
the Einstein frame. For an additional ideal fluid of energy-momentum
tensor $T^\mu_\nu={\rm diag}(-\delta \rho,\delta p,\delta p,\delta p)$
added to the dilaton
the energy density and pressure density appear in the Einstein--Friedmann
equations as
\beq
\delta\rho=6H^2-{1\over 2}\dot\varphi^2\,,\quad - \delta p=4\dot
H+6H^2+{1\over 2}\dot
\varphi^2\;.\label{rhop}
\eeq
 When both $\delta\rho$ and $\delta p$ vanish, we recover, of course, the  
tree-level solutions.
Inserting instead the corrected metric and dilaton backgrounds, we
easily obtain:
\beq
\delta \rho=-{4\over\sqrt 3}{D\over t^2}\,,\quad \delta p=0\;.
\eeq
Since $D$ is positive, see (\ref{result}) and (\ref{witht_P}), the energy
density is negative in the Einstein frame. The value $\delta p=0$
is  special to
 the background (\ref{solution}). We have checked that, for other power-law
 backgrounds
(e.g. for
radiation-like solutions) $\delta p$ does not vanish.

To translate these results to the string frame, we note that the
corresponding
energy and pressure densities are given by
\beq
e^\varphi\delta\rho_s&=&e^{-\varphi}\delta\rho=-3H_s^2+\dot{\overline\varphi}^2\nn
e^\varphi\delta p_s&=&e^{-\varphi}\delta p=2\dot
H_s-3H_s^2-2H_s\dot{\overline\varphi}-
\dot{\overline\varphi}^2+2\ddot{\overline\varphi}\;,
\eeq
so that, again, $\delta\rho_s<0$
and $\delta p_s=0$. Hence, the condition $\delta\rho_s+\delta p_s<0$
is clearly met,
see Fig. \ref{exit}.

We conclude that, from all points of view, the computed back-reaction
goes in the right direction to favour a change to the FRW branch and  
to avoid violation of the entropy bounds.

\vspace{.5cm}
\noindent{\Large\bf Appendix}

\vspace{.5cm}

\begin{figure}[htb]
\epsfxsize=5in
\bigskip
\centerline{\epsffile{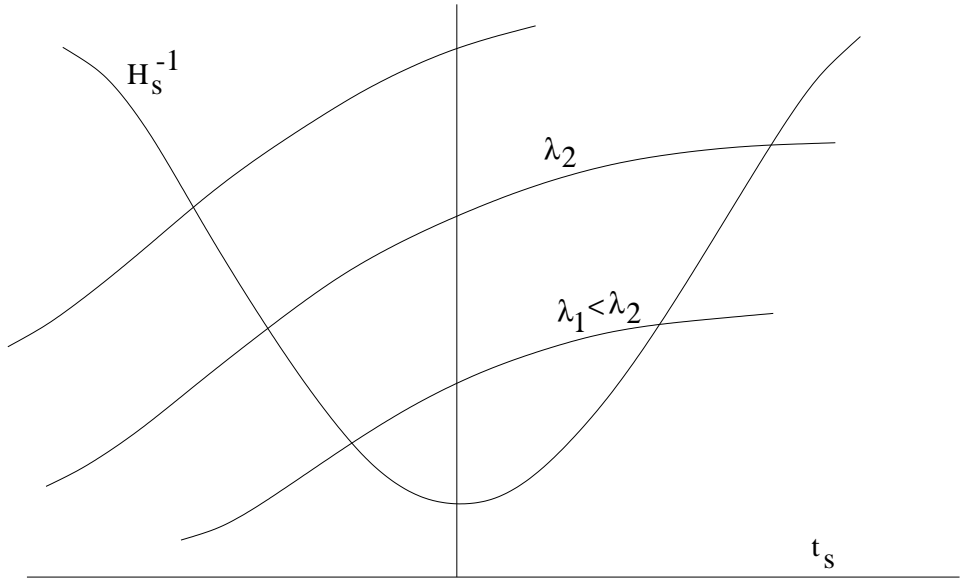}}
\caption{\it\baselineskip=12pt
Horizon crossings of various scales in PBB cosmology.}
\bigskip
\label{hubble}\end{figure}

\noindent
Here we compare our results with similar calculations done in other gauges,
such as the Iliopoulos
et al. gauge, which has been extensively used in the
recent literature to compute back-reaction to chaotic \cite{AW1} and  
power-law inflation \cite{AW2}. Specifically, in 
the latter case,
the authors found no strong effect (i.e. growing with time) at one loop.  
It might
not be so clear to the reader why the one-loop effect is so dominant in 
PBB inflation, which is also essentially power law. To clarify the  
situation it is
important to recall the basic distinction between PBB and  
standard power-law
inflation. In the PBB case the Hubble parameter grows with time, as $t\to 0^-$, 
and as a result higher frequency modes leave the horizon at higher values  
of the Hubble parameter, see Fig. \ref{hubble}.
The loop integrals are thus dominated by the latest value of the
Hubble parameter, in contrast with the case of standard power law,
where they are dominated by its initial value.
 It is also important to
keep in mind the two different time limits: in the case of
standard power law,
the limit is $t\to\infty$, whereas, in PBB it is, $t\to 0^-$.
This makes a crucial difference when deciding  
which are the dominant or subdominant effects.

As a check of gauge invariance, we feel it is worthwhile to redo 
the standard power-law calculation in the off-diagonal
gauge. It will provide a direct check of the results obtained  in the 
 gauge used in \cite{AW2}, as
well as help to better understand the differences between the two
cases in the same gauge. Let us
recall some of the basic formulas for standard power-law inflation. The
scale factor, Hubble
radius, etc., are given by ($s>1$):
\beq
\ln a_0=s\ln t_0=-{s\over s-1}\ln\eta\,,\quad
H_0=H_ia_0^{-1/s}={s\over t_0}\,,\quad
H_i=-{s\over s-1}{1\over\eta_i}\label{powers}
\eeq
and the dilaton is given by $\varphi_0=2\sqrt s\ln t_0$. For an arbitrary  
value of $s$, (\ref{powers}) describes
a background solution only in the presence of a potential
\beq
V(\varphi)=2H_i^2\left(3-{1\over s}\right)\exp\,[-\varphi/\sqrt  
s]\;.\label{potential}
\eeq
Note that for the special value $s=1/3$, corresponding to the PBB case, the  
potential
vanishes. To proceed as
before, in the off-diagonal gauge, we have to first find the quadratic  
constraints through
which the metric perturbations become related to the canonical variable
$\phi$. One finds
\beq
\Phi=-{1\over 2\sqrt s}\,\phi\,,\qquad\nabla^2B={1\over 2\sqrt s}\,\phi'\;.
\label{powercons}\eeq
The rest of the analysis is almost the same as before, except one  
should take
into account the extra three-point vertices coming from the potential
(\ref{potential}). Using the constraints and the gauge choice, the  
one-point  functions
turn out to have the following forms
\beq
A=0\,,\quad C=-{1\over\sqrt s}D\,,\quad D=-\Delta_0^{-1}\left(\delta-{1\over
\sqrt s}\gamma\right)\label{poweresult}
\eeq
where $\delta$ and $\gamma$ represent the same amputated diagrams as  
before.
After summing over the vertices, it turns out that the
leading effect in the one-point functions is of the form\thanks{For instance, 
for the particular case $s=2$, we found
$D=-{9\sqrt 2\over 512}\ln t_0$.} $D\sim\ln t_0$.
If we proceed, as in the case of PBB, to study the contribution of $C$ and $D$ 
to $\varphi$ and $H_{\rm one-loop}$, we find that the leading logarithms
disappear, irrespective of the
numerical factor in front of $\ln t_0$. The reason for the cancellation
is twofold: firstly, the relation (\ref{poweresult}) between $C$ and $D$
insures that the changes in $\varphi$ and $H$ are proportional. Secondly,
it is easy to show that for $C\sim\ln t_0$ the change in $H$ is subleading
since $\dot Ct/C\sim(1/\ln t)\to 0$.

Finally let us briefly point out another technical difference between the 
PBB and the standard power-law cases. 
In the PBB case the constraints relating $\Phi$ and $B$ to
$\phi$ have signs opposite to those in (\ref{powercons}) implying that
$C$ and $D$ have the same sign, see (\ref{one-point}). This is why their  
effects on $\varphi$ and $H_{\rm one-loop}$ add up, in contrast to  
the case of standard power law, where they compete with
each other. Furthermore, the leading
behaviour is no longer logarithmic, explaining why a large effect  
survives only in the PBB case.

%%%%%%%%%%%%%%%%%%%%%%%%%%%%%%%%%%%%%%%%%%%%%%%%%%%%%%%%%%%%%%%%
\subsection*{Acknowledgements}

We acknowledge useful discussions and correspondence with M. Gasperini
and R.P. Woodard. The work of A.G. was supported in part by the World 
Laboratory.
%%%%%%%%%%%%%%%%%%%%%%%%%%%%%%%%%%%%%%%%%%%%%%%%%%%%%%%%%%%%%%%%

\end{document}